# Magnetic Ordering Induced Giant Optical Property Change in Tetragonal BiFeO$_3$

Wen-Yi Tong,[1] Hang-Chen Ding,[1] Shi-jing Gong,[1] Xiangang Wan[2,3] and Chun-Gang Duan[1,*]

[1]Key Laboratory of Polar Materials and Devices, Ministry of Education, East China Normal University, Shanghai 200062, China

[2]National Laboratory of Solid State Microstructures and Department of Physics, Nanjing University, Nanjing 210093, China

[3]Collaborative Innovation Center of Advanced Microstructures, Nanjing University, Nanjing 210093, China

Magnetic ordering, as one of the most important characteristics in magnetic materials, could have significant influence on the band structure, spin dependent transport, and other important properties of materials. Its measurement, especially for the case of antiferromagnetic ordering, however, is generally difficult to be achieved. Here we demonstrate the feasibility of magnetic ordering detection using a noncontact and nondestructive optical method. Taking the compressive strained tetragonal BiFeO$_3$ (BFO) as an example and combining density functional theory calculations with the minimal one-band tight-binding models, we find that when BFO changes from C1-type antiferromagnetic (AFM) phase to G-type AFM phase, the top of valance band shifts from the Z point to Γ point, which makes the original direct band gap become indirect. This can be explained by the two-center Slater-Koster parameters using the Harrison approach. The impact of magnetic ordering on energy band dispersion dramatically changes the optical properties of tetragonal BFO. For the linear ones, the energy shift of the optical band gap could be as large as 0.4 eV. As for the nonlinear ones, the change is even larger. The second-harmonic generation coefficient $d_{33}$ of G-AFM becomes more than 13 times smaller than that of C1-type AFM case. Finally, we propose a practical way to distinguish the C1- and G-type AFM of BFO using the optical method, which might be of great importance in next-generation information storage technologies and widens the potential application of BFO to optical switch.

## I. INTRODUCTION

Magnetic materials have received intensive attention[1-10] in the last decades, as they generally contain rich mutual interactions of spin, charge, photon and lattice degree of freedom. They also have important applications in advanced technology such as magnetic read heads and magnetic memory cells, and many other new tools for magnetic imaging.[11-14] Despite extensive studies in the field of magnetic materials, the tools to obtain detailed information on the arrangement of magnetic moments in antiferromagnets are still limited, largely due to the fact that antiferromagnets have no or little macroscopic magnetization and therefore are insensitive to the external magnetic field.[15, 16] Considering the magnetic ordering often has important effect on the properties of the material,[17-24] a practical, inexpensive way to its measurement is in urgent demand, which is also important to both fundamental and technological developments in the field of magnetoelectronics and spintronics.

As an invaluable experimental probe, the optical spectroscopy is of great scientific and practical interest due to its advantage in providing complementary information on crystallographic, electronic and magnetic properties and studying the coexistence and interactions of magnetic and electric order.[25, 26] With the rapid development of spintronics and its cousin valleytronics, the optical spectroscopy is regarded as a promising candidate to probe and manipulate the spin and valley degrees of freedom.[27-31] Compared with neutron scattering method, which is regarded as the most powerful and versatile experimental technique in obtaining truly three-dimensional pictures of magnetic structures in various materials,[32] optical spectroscopy techniques possess the advantage of relative small samples, simplicity of source preparation, rapidity of measurement in a non-contact way. Since the arrangement of spins has great influence on the interplay between matter and electromagnetic radiation, which will directly be reflected in optical properties of materials, is it possible to determine the magnetic ordering using optical methods?

In this work, we focus our study on the perovskite oxide bismuth ferrite ($BiFeO_3$, BFO), which is the very rare single-phase multiferroic material simultaneously with antiferromagnetism and robust ferroelectricity well above room temperature,[33, 34] and has attracted extensive research.[35-44] Using first-principles calculations, we investigate the complex dielectric functions and second-harmonic generation (SHG) coefficients of expitaxially stabilized BFO in tetragonal phase with giant $c/a$ ratio and find sizable differences for both linear and nonlinear optical properties between C1- and G-antiferromagnetic (AFM) ordering. The predicted modification of optical properties induced by magnetic ordering change reveals the possibility of using a noncontact and nondestructive optical method to distinguish the C1- and G-type AFM of BFO, which is currently hard to be handled by a traditional neutron diffraction measurement because of the strong diamagnetic response of substrate materials.[45]

## II. COMPUTATIONAL METHOD

The calculations are performed within density-functional theory (DFT) using the accurate full-potential projector augmented wave (PAW) method, as implemented in the Vienna *ab initio* Simulation Package (VASP).[46] The exchange-correlation potential is treated in Perdew-Burke-Ernzerhof (PBE) form[47] of the generalized gradient approximation (GGA) with a kinetic-energy cutoff of 500 eV. A 7×7×7 and 11×11×11 Monkhorst-Pack $k$-point mesh centered at Γ are respectively adopted in the self-consistent and optical calculations. A primitive cell, containing two BFO formula units (10 atoms) is performed to describe the magnetic structures of BFO. Due to the different shapes of the primitive lattice between C1- and G-type AFM cases, all the coordinates mentioned below are described in real or reciprocal space of C1-AFM. The convergence criterion for the electronic energy is $10^{-6}$ eV and the structures are relaxed until the Hellmann-Feynman forces on each atoms become less than 1 meV/Å. The GGA + $U$ method[48] with the effective Hubbard constant $U_{eff} = U - J = 2.0$ eV[44] is chosen to give a better description of the partially filled and strongly correlated localized Fe-3$d$ electrons. We also check that our results are robust within the other widely used $U_{eff}$'s as 4.0 eV and 6.0 eV.

The optical calculations are performed using our own code OPTICPACK, which has been successfully applied to study the optical properties of borate crystals[49, 50] and other multiferroic materials.[51] For the linear case, the imaginary part of the complex dielectric function $\varepsilon_2$ is calculated using the following relations:

$$\varepsilon_2(E) = \frac{8\pi^2}{\Omega} \sum_{\vec{k}} W_{\vec{k}} \sum_{v,c} |p_{cv}|^2 \frac{\delta(E_c - E_v - E)}{(E_c - E_v)^2}, \quad (1)$$

where $E$ is the photon energy, $\Omega$ is the cell volume, $p_{cv}$ is the electron momentum matrix element between the valance band (VB) states ($v$) and the conduction band (CB) states ($c$). The integral over the $k$ space has been replaced by a summation over special $k$ points with corresponding weighting factor $W_{\vec{k}}$. The second summation includes $v$ and $c$ states, based on the reasonable assumption that the VB is fully occupied, while the CB is empty.

For the nonlinear case, we only consider the SHG susceptibility $\chi^{(2)}(2\omega; \omega, \omega)$ in the static limit, which can be written as:

$$\chi^{(2)}_{ijk}(0) = \frac{1}{\Omega}[\sum_{vv'c} P(ijk) \text{Im}(p^i_{v'v} p^j_{vc} p^k_{cv'})(\frac{1}{E^3_{cv}E^2_{cv'}} + \frac{2}{E^4_{cv}E_{cv'}}) \quad (2)$$
$$+ \sum_{vcc'} P(ijk) \text{Im}(p^i_{cv} p^j_{vc'} p^k_{c'c})(\frac{1}{E^3_{cv}E^2_{c'v}} + \frac{2}{E^4_{cv}E_{c'v}})],$$

in which $P(ijk)$ denotes full permutation over Cartesian components $i$, $j$, $k$ and explicitly shows Kleinman symmetry[52] of the SHG coefficients. The first and second summations represent virtual-hole and virtual-electron process respectively.

### III. RESULTS AND DISCUSSION

A tetragonal BFO with giant $c/a$ ratio, i.e. the so-called super-tetragonal structure,[53] is more energetically favorable under compressive strains with an "out of plane" ferroelectric polarization.[38] There exists a transition between C1- and G-type AFM ordering for tetragonal BFO under the influence of in-plane strain.[17, 54] As a result, here we establish the [001] polarized BFO in tetragonal perovskite structure ($C_{4v}$, $P4mm$) with the lattice constant of 3.905 Å, around which is the transformation point for the two AFM phases. The optimized $c/a$ ratio is 1.197.

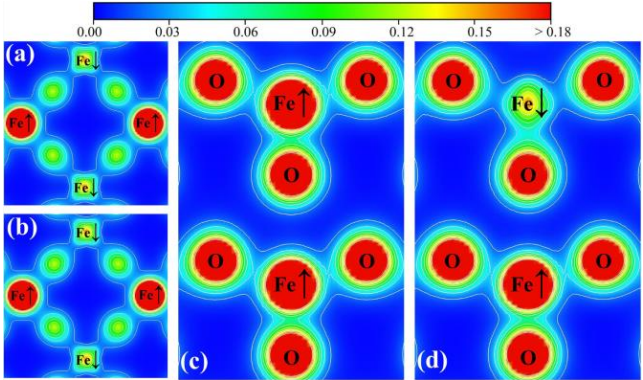

FIG. 1. Majority-spin charge densities in the (001) plane for (a) C1-type and (b) G-type AFM cases. (c) and (d) are those in the (110) plane for C1- and G-AFM, respectively. The up and down arrows on Fe atoms indicate the spin orientations.

To explore the effect of magnetic ordering on the electronic structure, the spin-dependent charge density is needed to be inspected. As an example, in Fig. 1, we display the majority-spin ones. Noted that on account of AFM ordering for both C1 and G states, the minority-spin charge density plots can be directly obtained by shifting the corresponding majority-spin ones by the in-plane lattice constant $a$ along the $x$- or equivalently $y$-axis. Since all the eight operations in point group $C_{4v}$ are $z$-irrelevant, the same intraplane couplings of C1 and G cases signify the same magnetic point group, i.e. $C_{2v}$. In $xy$ plane, the magnetic coupling of neighboring magnetic ions in both states are AFM, which leads to essentially the same charge density distribution. On the contrary, when we consider the $z$ direction, there exists significant difference in the spin-dependent charge density of Fe atoms. In C1-AFM, whose nearest interlayer coupling is ferromagnetic (FM), as shown in Fig. 1(c), the density changes continuously along the $z$ direction in the spin-up channel, and so does the spin-down one. For G-AFM, however, the nearest AFM interlayer coupling leads to the discontinuous charge density distribution, which inevitably affect the spin-transport properties of the BFO film.

The spin arrangement along the $z$-direction not only have influence on transport behaviors, but also will affect the band structure. The energy spectra $E_n(\boldsymbol{k})$ near the Fermi level are displayed in Fig. 2. We emphasize that to make a straightforward comparison between the band structures of two AFM phases, the DFT-calculated bands along high-symmetry directions are identical in the $k$-space, although their first Brillouin zones (FBZ) are indeed different. In both C1- and G-type AFM orderings, the ingredients of electronic states are quite similar. The presented VB close to the Fermi level is predominantly comprised of Fe-$e_g$ and O-$2p$ states. The band mainly from the Bi-$p_z$ orbital is highly dispersive along $\Gamma$-Z and Z-A directions. The flat band lying around 1.7 eV above the Fermi level is primarily of the empty Fe-$3d_{xy}$ character. Nevertheless, the dispersions of energy bands of the two magnetic cases are significantly different. For C1-type AFM, the top edge of VB, as well as the bottom of CB are located at the $Z$ (0, 0, 0.5) point. The band gap ($E_g$) is direct, about 1.12 eV. For G-type AFM, although the lowest unoccupied molecular orbital (LUMO) is still at the Z point, the highest occupied molecular orbital (HOMO) shifts to the center of the FBZ. Its $E_g$ is indirect with the value of 1.03 eV. The direct gap $E_g$ is much larger, i.e. 1.51 eV at the Z point. When we carefully analyze the band features, especially along the $\Gamma$ - Z symmetry line, it is intriguing to point out that due to the occupancy of the Fe-$3d$ states, the extreme points of bands would be shifted during the transformation of magnetic ordering. As an example, for C1-AFM, the maximum and minimum of the $d_{z^2}$ and $d_{xy}$ bands are located at Z and $\Gamma$, respectively. For G-AFM, they are just reversed. However, it would not happen for the band dominated by the $p$ electrons of Bi atoms. In order to accurately capture the physics about the impact of Fe-3$d$ orbitals on the energy band dispersions between two AFM states, particularly the shift of the HOMO, tight-binding (TB) model calculations are then carried out.

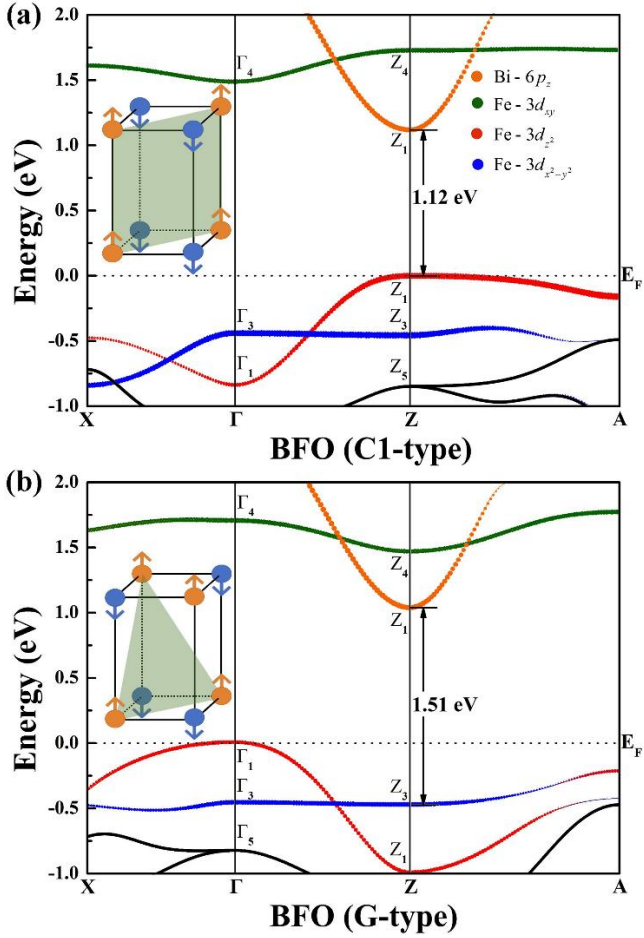

FIG. 2. Band structures of tetragonal BFO under (a) C1-type AFM ordering, and (b) G-type AFM ordering. The spin arrangements of each ordering are shown as insets. The radius of dots represent the weight of Fe-3$d$ and Bi-6$p$ characters for various subbands. The Fermi level $E_F$ is set to zero.

As our purpose is to show the band difference between different magnetic configurations, we adopt an effective, iron-only one-band TB model. This is based on the fact that except for the mixing between $d_{z^2}$ and $d_{x^2-y^2}$ orbitals along Γ-A direction in BFO under G-AFM ordering, the $d_{xy}$, $d_{z^2}$ and $d_{x^2-y^2}$ states are basically independent, as shown in Fig. 2. Since the electrons hopping between Fe ions generally needs the O-2$p$ orbitals as medium, the iron-oxygen hybridization has already been included in the effective hopping term of our iron-only model. Particularly, the minimal TB model involving Fe-$d_{z^2}$ orbital is used to reveal the behavior of the top edge of VB.

To be specific, the interatomic matrix elements of the non-magnetic Hamiltonian can be written in terms of on-site energies $\varepsilon_{\alpha\alpha}$ and effective hopping integrals $t_{b\beta,a\alpha}$ as follows:

$$<b\beta \mathbf{k}|H_{TB}|a\alpha \mathbf{k}> = \delta_{b\beta,a\alpha}\varepsilon_{\alpha\alpha} + \sum_n e^{i\mathbf{k}\cdot\mathbf{R}_n} t_{b\beta,a\alpha} \quad (3)$$

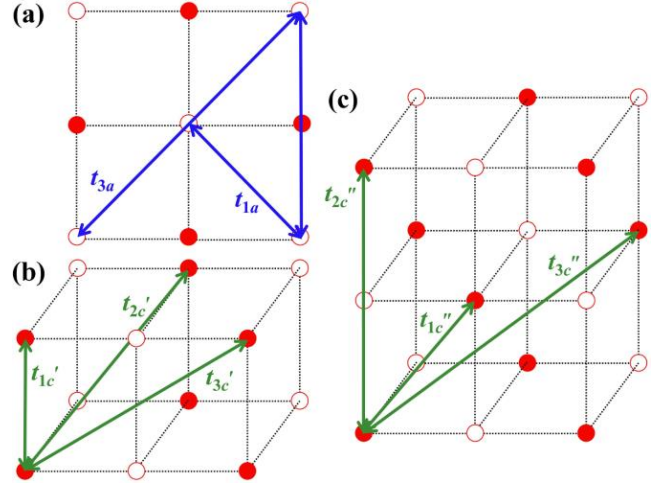

FIG. 3. The FM interactions between the Fe sites in BFO. (a) is the projection view on the $xy$ plane, (b) and (c) are the off-$xy$-plane interactions for C1- and G-AFM. The solid and hollow circles represent the spin-up and spin-down Fe atoms, respectively.

Here $\mathbf{k}$ is the Bloch wave vector. $a$ ($b$) and $\alpha$ ($\beta$) are the atom index and the atomic orbital quantum number, respectively. In order to take the magnetic coupling into account, Hilbert space should be doubled by including the spin degree of freedom ($\sigma$), i.e. $|a\alpha \mathbf{k}> \rightarrow |a\alpha\sigma \mathbf{k}>$. Then the effective hopping integral between any two ions depends on $\cos(\theta/2)$,[55] where $\theta$ is the angle between their spins. Consequently, i.e. only when the coupling is FM, there exists hopping between the two magnetic ions and the interaction of them should then be invoked in the Hamiltonian. Considering that for tetragonal BFO under C1- and G-type AFM ordering, the magnetic coupling of Fe ions in the $xy$ plane are exactly the same, we write the TB Hamiltonian referring to the Fe-3$d$ orbitals as:

$$H_{C1}(\mathbf{k}) = H_{xy}(\mathbf{k}) + t_{1c}' c_z(c) + t_{2c}' c_z(c)[c_x(a) + c_y(b)]$$
$$+ t_{3c}' c_z(c) c_x(a) c_y(b),$$

$$H_G(\mathbf{k}) = H_{xy}(\mathbf{k}) + t_{1c}'' c_z(c) c_x(\frac{a}{2}) c_y(\frac{b}{2}) + t_{2c}'' c_z(2c) \quad (4)$$
$$+ t_{3c}'' c_z(c)[c_x(\frac{a}{2}) c_y(\frac{3b}{2}) + c_x(\frac{3a}{2}) c_y(\frac{b}{2})],$$

where

$$H_{xy}(\mathbf{k}) = \varepsilon_d + t_{1a}[c_x(a) + c_y(b)] \\ + t_{2a}c_x(a)c_y(b) + t_{3a}[c_x(2a) + c_y(2b)] \quad (5)$$

with $c_i(R_i) = 2\cos(k_i R_i)$ ($i = x, y, z$) and $R_i$ is the lattice constant along the $x$, $y$ or $z$ direction. $\varepsilon_d$ refers to the single particle energy. We consider the intralayer and interlayer hopping between first, second and third FM neighbors for C1- and G-AFM, accordingly. Test calculations show that higher order hopping terms are negligible and therefore are not counted. The interaction pathways corresponding to the effective hopping parameters are shown in Fig. 3.

A straightforward analysis show that for both magnetic configuration the band reaches its extreme point at $\Gamma$ and Z. The energy differences between them are given by:

$$\Delta E_{C1} = H_{C1}^\Gamma - H_{C1}^Z = 4t_{1c}' + 16t_{2c}' + 16t_{3c}', \\ \Delta E_G = H_G^\Gamma - H_G^Z = 16t_{1c}'' + 32t_{3c}''. \quad (6)$$

Obviously, locations of the maximum and minimum for the band are determined by the interlayer hopping parameters.

TABLE I. Values of effective tight-binding parameters for the $d_{z^2}$ band (in meV).

| $d_{z^2}$ | C1-AFM | G-AFM |
|---|---|---|
| $\varepsilon_d$ | -259.18 | -259.18 |
| $t_{1a}$ | -34.56 | -34.56 |
| $t_{2a}$ | 1.33 | 1.33 |
| $t_{3a}$ | -7.49 | -7.49 |
| $t_{1c}'$ | -97.14 | - |
| $t_{2c}'$ | -27.32 | - |
| $t_{3c}'$ | -1.15 | - |
| $t_{1c}''$ | - | 63.72 |
| $t_{2c}''$ | - | -36.29 |
| $t_{3c}''$ | - | -0.94 |

As listed in Table I, effective hopping integrals of the one-band TB model involving Fe-$d_{z^2}$ orbitals are obtained by fitting to the first-principles results. For C1-type AFM case, the negative parameter $t_{1c}'$ causes that the subband has higher energy in the $\Gamma$ point than in the Z point. For G-AFM, however, the interlayer nearest hopping parameter $t_{1c}''$ is positive and then leads to the opposite results. The two-center Slater-Koster parameters[56] using the Harrison approach[57] are applied to explain the sign difference between them. For $d_{z^2}$ orbitals, the hopping term is a function of the direction cosines, $l$, $m$, $n$ of the vector $x$, $y$ and $z$, as shown below.

$$t = [n^2 - \frac{1}{2}(l^2 + m^2)]^2 V_{dd\sigma} + 3n^2(l^2 + m^2)V_{dd\pi} \\ + \frac{3}{4}(l^2 + m^2)^2 V_{dd\delta}. \quad (7)$$

The three $dd$ matrix elements are all obtained from

$$V_{ddm} = \eta_{ddm} \frac{\hbar^2 r_d^3}{m d^5}, \quad (8)$$

where $r_d$ is a length that is characteristic of each element and $d$ is the distance between atoms. $\hbar$ and $m$ are reduced Planck constant and electron mass, respectively. Then the terms $t_{1c}'$ and $t_{1c}''$ can be written as:

$$t_{1c}' = \eta_{dd\sigma} \frac{\hbar^2 r_d^3}{m d^5} = \frac{1}{(c/a)^5} \eta_{dd\sigma} \frac{\hbar^2 r_d^3}{m a^5},$$

$$t_{1c}'' = [(\frac{2(c/a)^2 - 1}{2(1+(c/a)^2)})^2 \eta_{dd\sigma} + \frac{3(c/a)^2}{(1+(c/a)^2)^2} \eta_{dd\pi} \quad (9) \\ + \frac{3}{4(1+(c/a)^2)^2} \eta_{dd\delta}] \frac{\hbar^2 r_d^3}{m d^5}$$

$$= \frac{((c/a)^2 - 1/2)^2 \eta_{dd\sigma} + 3(c/a)^2 \eta_{dd\pi} + 3/4 \eta_{dd\delta}}{(\sqrt{1+(c/a)^2})^9} \frac{\hbar^2 r_d^3}{m a^5}.$$

For the term $t_{1c}'$, since the interaction path is just along the $z$ direction, only $\sigma$-orbital combination exists between the Fe atomic orbitals. Same signs of the wave functions makes the coefficient $\eta_{dd\sigma}$ negative, and then gives rise to the negative $t_{1c}'$. While, for the term $t_{1c}''$, there are angular momentum around all the three vectors, which triggers the coexistence of $\sigma$, $\pi$ and $\delta$ orbital combinations. The coefficient $\eta_{dd\delta}$ is negligible compared with $\eta_{dd\sigma}$ and $\eta_{dd\pi}$. The $\pi$-orbital combination has factor $3(c/a)^2$, which is almost 5 times larger than the factor $((c/a)^2-1/2)^2$ for $\sigma$-orbital combination. Therefore, the $\sigma$-orbital combination is suppressed by the $\pi$-orbital combination. The positive coefficient $\eta_{dd\pi}$ makes the term $t_{1c}''$ positive. We also examine other effective hopping parameters in Table I. The signs of them are all consistent with that of the two-center Slater-Koster parameters using the Harrison approach.

We emphasize that the effective hopping parameters $t_{1c}'$ and $t_{1c}''$ are the most important fitting parameters to explain the behavior of the $d_{z^2}$ band. Sign difference between them directly controls the location of the HOMO. When the magnetic ordering of tetragonal BFO changes from C1- to G-type, the interlayer nearest hopping term varies from the negative $t_{1c}'$ to the positive $t_{1c}''$, directly triggering the top of VB shifts from the Z point to the center of the FBZ.

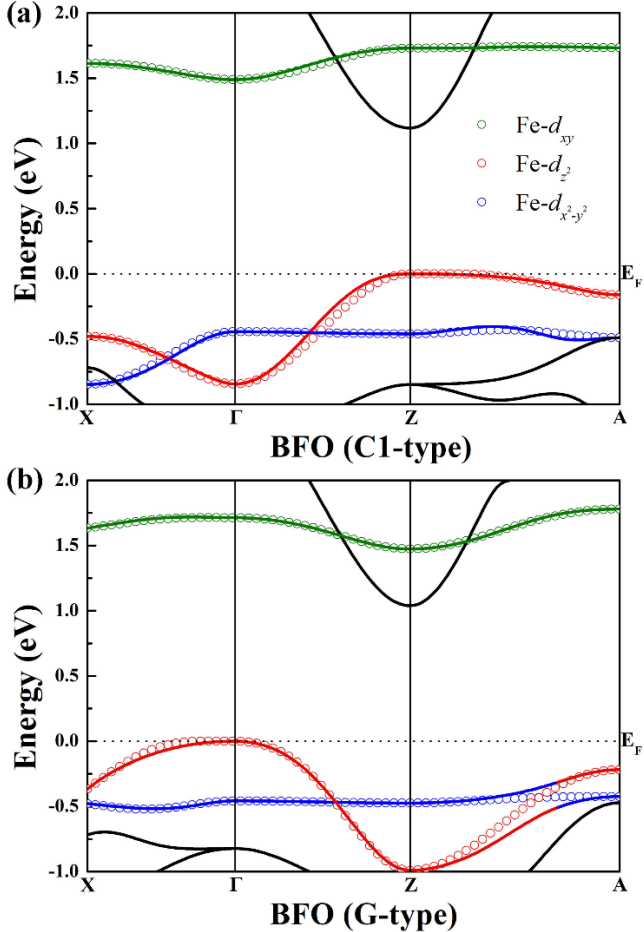

FIG. 4. Comparison of the model TB bands (open circles) and the corresponding first-principles results (solid lines). The green, red and blue circles represent the $d_{xy}$, $d_{z^2}$ and $d_{x^2-y^2}$ characters of Fe-3d, accordingly. The values of TB effective parameters are obtained via an optimal fit to the first-principles bands.

In addition to the $d_{z^2}$ orbital we discussed above, the minimal TB model involving $d_{x^2-y^2}$ and $d_{xy}$ orbitals of Fe-3d are also taken into account. Fig. 4 compares the model TB bands (open circles) with the corresponding DFT results (solid lines). Due to the hybridization between $d_{z^2}$ and $d_{x^2-y^2}$ states, which is not considered in our simple one-band TB models, there exists mismatch in bands dispersion along the Z–A symmetry line for G-type BFO. Even so, the agreement between the TB models and the first principles calculations is excellent, which warrants the validity of the TB models we adopted here.

According to previous discussions, the influence of magnetic ordering on band structures primarily reflects on the shift of the HOMO, the reason of which has been explained by the simple yet convincing TB models. As a result, in the process of phase transformation from C1- to G-AFM, the direct band gap becomes indirect. It is interesting to point out that, in contrast to the decrease of band gap (~ 0.09 eV), the enhancement of direct band gap reaches up to 0.40 eV. It, apparently, will greatly affect the optical properties of BFO.

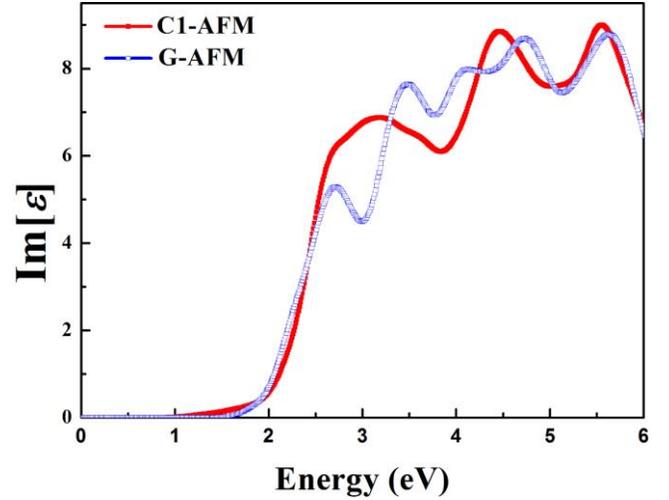

FIG. 5. The imaginary parts of complex dielectric function $\varepsilon_2$ in the near infrared to the near ultraviolet range of tetragonal BFO for the case of C1-AFM ordering (red solid cube line), and G-AFM ordering (blue hollow cube line).

For the linear optical properties, the imaginary parts of complex dielectric functions $\varepsilon_2$ are calculated and shown in Fig. 5. Note that as a second-order process with a relatively small transition probability, the indirect transition process would happen only if phonons participate in and is therefore ignored here. The interband transitions begin at 1.12 eV and 1.52 eV for C1-AFM and G-AFM, respectively, in good agreement with the direct $E_g$ gained by band structures. For C1-type AFM case, owing to the selection rules, the transition from HOMO to LOMO in the Z point is allowed when the incident radiation is polarized in $z$-direction. The $E_g^{opt}$ exactly equals to the direct $E_g$. However, for G-type AFM case, for any polarization direction of incident light, the electric-dipole transition related to the direct $E_g$ in the Z point is forbidden. Nevertheless, it is allowed when the $k$-point moves to the Z–A symmetry line. This suggests that $E_g^{opt}$ in this case should be slightly larger than the already increased direct $E_g$. Consequently, as shown clearly in Fig. 5, when the magnetic ordering varies from C1- to G-type AFM, the optical band gap ($E_g^{opt}$) of tetragonal BFO exhibits a blue shift of approximately 0.4 ~ 0.5 eV, which is large enough to be easily distinguished in experimental optical spectra.

TABLE II. SHG coefficients $d_{ij}$ (pm/V) in the static limit for the tetragonal BFO with C1- and G-AFM ordering.

| SHG components | C1-AFM | G-AFM |
| --- | --- | --- |
| $d_{15}$ | 59.550 | 57.898 |
| $d_{33}$ | -78.905 | -5.900 |

Besides the complex dielectric functions, the nonlinear optical properties, especially in the static limit, are more sensitive to the change of the direct $E_g$. The SHG susceptibilities in static limit are listed in Table II. Due to the symmetry of tetragonal BFO with the point group of C$_{4v}$, there are five independent SHG components, i.e. $d_{15} = d_{24}$, $d_{31} = d_{32}$ and $d_{33}$. The so-called Kleinman symmetry[52] demands $d_{15} = d_{31}$ in the static limit. Though the $d_{15}$ component between the two kinds of AFM ordering is about the same, the absolute value of $d_{33}$ for C1-type AFM phase is more than 13 times larger than that of G-type AFM phase. The considerable difference with an order of magnitude reveals its utility to measure the magnetic ordering.

Finally we propose a practical device to distinguish the C1- and G-type AFM of BFO using optical methods. As shown in Fig. 6, the tetragonal BFO under the [001] polarization direction with the lattice constant of 3.905Å, around which the C1-G magnetic transition will occur, is placed onto a PMN-PT ([Pb(Mg$_{1/3}$Nb$_{2/3}$)O$_3$]$_{0.72}$[PbTiO$_3$]$_{0.28}$) actuator via gluelike PMMA (polymethyl methacrylate) layer. The biaxial strain is provided by the PMN-PT substrate sandwiched between transparent In$_2$O$_3$ thin films acting as electrodes. A bias voltage $V$ applied to the PMN-PT results in an out-of-plane electric field $E$, which leads to an in-plane strain $\varepsilon$. The PMN-PT is electrically poled so that $V > 0$ ($< 0$) corresponds to in-plane compressive (tensile) strain. Several experimental works[58,59] demonstrated the reversibility of the strain tuning technique. In order not to affect the transmission of light, the materials serving as actuator, glue and electrode are carefully selected to be highly near-infrared transparent. A laser beam with the energy $h\nu = 1.3$ eV is incident to the tetragonal BFO. Note that as DFT calculation generally will underestimate the energy gap, the realistic photon energy could be larger. For $V > 0$, the tetragonal BFO experiences an in-plane compression and is in the C1-AFM state. Under this circumstance, electrons will absorb the photons and jump to the excited state, i.e. the film is opaque. By applying a negative voltage $V < 0$ to the substrate, however, the BFO film transforms to G-type AFM. Since now the energy gap is larger than the photon energy, the film then becomes transparent, and the incident light could be detected by the IR detector beneath.

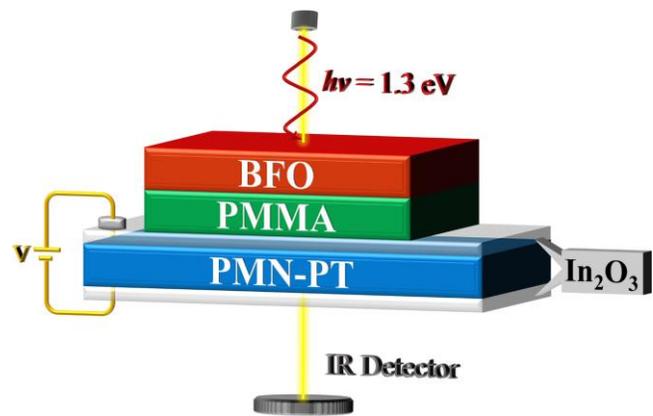

FIG. 6. Proposed new device utilizing the magnetic ordering induced giant optical properties change.

In the above device, the magnetic orderings can be controlled by the sign of the bias voltage $V$. And the magnetic states can be easily "read out" utilizing the IR detector to monitor the transmitted light. Obviously, it might also be used as the electrically writing and optically reading memory devices.

## IV. CONCLUSION

Using first-principles calculations, we found that the different spin arrangements in tetragonal BFO have significant influence on the dispersion of energy bands, especially the location of the HOMO, which has been explained by the minimal one-band TB models. When the magnetic ordering of tetragonal BFO varies from C1- to G-type, sign difference between the interlayer nearest hopping terms directly triggers the shift of HOMO from the Z point to Γ point. Consequently, the energy gap changes from direct to indirect one and leads to drastic modification to the optical properties of tetragonal BFO. For the linear ones, the enhancement of $E_g^{opt}$ can be as high as 0.4 eV. When it comes to the nonlinear ones, the change of the SHG coefficient $d_{33}$ becomes more than 13 times. Obviously, the difference of optical properties are considerable enough to distinguish the C1-type and G-type AFM of tetragonal BFO experimentally. We therefore hope that our theoretical predictions will stimulate experimental studies of magnetic ordering measurements using a noncontact and nondestructive optical method.

## AUTHOR INFORMATION

Email: cgduan@clpm.ecnu.edu.cn

Notes
The authors declare no competing financial interest.

## ACKNOWLEDGMENTS

This work was supported by the 973 Programs (No. 2014CB921104, 2013CB922301), the NSF of China (No. 61125403, 11374137), NSF of Shanghai (No. 14ZR1412700), Program of Shanghai Subject Chief Scientist. Computations were performed at the ECNU computing center.

## ■ REFERENCES

[1] I. Žutić, J. Fabian, and S. Das Sarma, Rev. Mod. Phys. **76**, 323 (2004).
[2] M. Fiebig, T. Lottermoser, T. Lonkai, A. Goltsev, and R. Pisarev, J. Magn. Magn. Mater. **290**, 883 (2005).
[3] C.-W. Nan, M. I. Bichurin, S. Dong, D. Viehland, and G. Srinivasan, J. Appl. Phys. **103**, 031101 (2008).
[4] K. F. Wang, J. M. Liu, and Z. F. Ren, Adv. Phys. **58**, 321 (2009).
[5] O. Gutfleisch, M. A. Willard, E. Bruck, C. H. Chen, S. G. Sankar, and J. P. Liu, Adv. Mater. **23**, 821 (2011).
[6] W.-G. Wang, M. Li, S. Hageman, and C. L. Chien, Nat. Mater. **11**, 64 (2012).
[7] S.-J. Gong, H.-C. Ding, W.-J. Zhu, C.-G. Duan, Z. Zhu, and J. Chu, Sci. China Phys. Mech. Astron. **56**, 232 (2013).
[8] X. Huang and S. Dong, Mod. Phys. Lett. B **28** (2014).
[9] R. O. Cherifi, et al., Nat. Mater. **13**, 345 (2014).
[10] F. Matsukura, Y. Tokura, and H. Ohno, Nat. Nanotech. **10**, 209 (2015).
[11] S. Steinert, et al., Nat. Commun. **4** (2013).
[12] J. J. Turner, et al., Phys. Rev. Lett. **107** (2011).
[13] N. Rougemaille and A. K. Schmid, Eur Phys J-Appl Phys **50** (2010).
[14] M. R. Freeman and B. C. Choi, Science **294**, 1484 (2001).
[15] B.-Y. Wang, J.-Y. Hong, K.-H. O. Yang, Y.-L. Chan, D.-H. Wei, H.-J. Lin, and M.-T. Lin, Phys. Rev. Lett. **110**, 117203 (2013).
[16] F. Nolting, et al., Nature **405**, 767 (2000).
[17] H.-C. Ding and C.-G. Duan, Europhys. Lett. **97**, 57007 (2012).
[18] H. J. Xiang, E. J. Kan, Y. Zhang, M. H. Whangbo, and X. G. Gong, Phys. Rev. Lett. **107**, 157202 (2011).
[19] J. M. Rondinelli, M. Stengel, and N. A. Spaldin, Nat. Nanotech. **3**, 46 (2008).
[20] C.-G. Duan, J. P. Velev, R. F. Sabirianov, Z. Zhu, J. Chu, S. S. Jaswal, and E. Y. Tsymbal, Phys. Rev. Lett. **101**, 137201 (2008).
[21] Y.-H. Chu, et al., Nat. Mater. **7**, 478 (2008).
[22] M. Weisheit, S. Fahler, A. Marty, Y. Souche, C. Poinsignon, and D. Givord, Science **315**, 349 (2007).
[23] C.-G. Duan, S. S. Jaswal, and E. Y. Tsymbal, Phys. Rev. Lett. **97**, 047201 (2006).
[24] H. Ohno, D. Chiba, F. Matsukura, T. Omiya, E. Abe, T. Dietl, Y. Ohno, and K. Ohtani, Nature **408**, 944 (2000).
[25] M. Fiebig, V. V. Pavlov, and R. V. Pisarev, J Opt Soc Am B **22**, 96 (2005).
[26] M. Fiebig, T. Lottermoser, D. Frohlich, A. V. Goltsev, and R. V. Pisarev, Nature **419**, 818 (2002).
[27] D. MacNeill, C. Heikes, K. F. Mak, Z. Anderson, A. Kormányos, V. Zólyomi, J. Park, and D. C. Ralph, Phys. Rev. Lett. **114**, 037401 (2015).
[28] A. V. Nalitov, G. Malpuech, H. Terças, and D. D. Solnyshkov, Phys. Rev. Lett. **114**, 026803 (2015).
[29] M. Gibertini, F. M. D. Pellegrino, N. Marzari, and M. Polini, Phys. Rev. B **90**, 245411 (2014).
[30] H. Zeng, J. Dai, W. Yao, D. Xiao, and X. Cui, Nat Nano **7**, 490 (2012).
[31] K. F. Mak, K. He, J. Shan, and T. F. Heinz, Nat Nano **7**, 494 (2012).
[32] J. Rodríguez-Carvajal, Physica B: Condensed Matter **192**, 55 (1993).
[33] G. Catalan and J. F. Scott, Adv. Mater. **21**, 2463 (2009).
[34] J. R. Teague, R. Gerson, and W. J. James, Solid State Commun. **8**, 1073 (1970).
[35] D. Sando, A. Barthelemy, and M. Bibes, J. Phys.: Condens. Matter **26**, 473201 (24 pp.) (2014).
[36] H. Yamada, et al., ACS Nano **7**, 5385 (2013).
[37] H. Liu, P. Yang, K. Yao, K. P. Ong, P. Wu, and J. Wang, Adv. Funct. Mater. **22**, 937 (2012).
[38] J. X. Zhang, et al., Phys. Rev. Lett. **107**, 147602 (2011).
[39] N. A. Spaldin, S.-W. Cheong, and R. Ramesh, Phys. Today **63**, 38 (2010).
[40] I. C. Infante, et al., Phys. Rev. Lett. **105** (2010).
[41] S. Ju, T. Y. Cai, and G. Y. Guo, J. Chem. Phys. **130** (2009).
[42] R. J. Zeches, et al., Science **326**, 977 (2009).
[43] J. B. Neaton, C. Ederer, U. V. Waghmare, N. A. Spaldin, and K. M. Rabe, Phys. Rev. B **71**, 014113 (2005).
[44] C. Ederer and N. A. Spaldin, Phys. Rev. Lett. **95**, 257601 (2005).
[45] G. J. MacDougall, H. M. Christen, W. Siemons, M. D. Biegalski, J. L. Zarestky, S. Liang, E. Dagotto, and S. E. Nagler, Phys. Rev. B **85** (2012).
[46] G. Kresse and J. Furthmüller, Comput. Mater. Sci. **6**, 15 (1996).
[47] J. P. Perdew, K. Burke, and M. Ernzerhof, Phys. Rev. Lett. **77**, 3865 (1996).
[48] S. L. Dudarev, G. A. Botton, S. Y. Savrasov, C. J. Humphreys, and A. P. Sutton, Phys. Rev. B **57**, 1505 (1998).
[49] C.-G. Duan, J. Li, Z.-Q. Gu, and D.-S. Wang, Phys. Rev. B **60**, 9435 (1999).
[50] C.-G. Duan, J. Li, Z.-Q. Gu, and D.-S. Wang, Phys. Rev. B **59**, 369 (1999).
[51] W.-Y. Tong, H.-C. Ding, Y.-C. Gao, S.-J. Gong, X. Wan, and C.-G. Duan, Phys. Rev. B **89**, 064404 (2014).
[52] D. A. Kleinman, Phys. Rev. **126**, 1977 (1962).
[53] H. Bea, et al., Phys. Rev. Lett. **102**, 217603 (2009).
[54] A. J. Hatt, N. A. Spaldin, and C. Ederer, Phys. Rev. B **81**, 054109 (2010).
[55] M. Baublitz, C. Lane, H. Lin, H. Hafiz, R. S. Markiewicz, B. Barbiellini, Z. Sun, D. S. Dessau, and A. Bansil, Sci. Rep. **4** (2014).
[56] J. C. Slater and G. F. Koster, Phys. Rev. **94**, 1498 (1954).
[57] W. A. Harrison, Electronic Structure and the Properties of Solids, San Francisco, 1980).
[58] F. Ding, H. Ji, Y. Chen, A. Herklotz, K. Doerr, Y. Mei, A. Rastelli, and O. G. Schmidt, Nano Lett. **10**, 3453 (2010).
[59] F. Ding, et al., Phys. Rev. Lett. **104** (2010).